\documentstyle[psfig]{mn}

\title[Optical and UV observations of a strong flare in LQ Hya]
{Optical and UV observations of a strong flare in 
the young, single K2-dwarf LQ Hya
\thanks{Based on observations made with the William Herschel 
operated on the island of La Palma by the
Royal Greenwich Observatory at the Spanish Observatorio 
del Roque de Los Muchachos of the 
Instituto de Astrof\'{\i}sica de Canarias,
and on data obtained with the International Ultraviolet Explorer}
}
\author[D. Montes et al.]
  {D.~Montes,$^{1,2}$ S.H.~Saar,$^3$ A. Collier Cameron,$^4$ Y.C. Unruh$^5$\\  
  $^1$Departamento de Astrof\'{\i}sica, Facultad de F\'{\i}sicas,
       Universidad Complutense de Madrid, E-28040 Madrid, Spain\\
  $^2$Pennsylvania State University, Department of Astronomy \& Astrophysics,
       525 Davey Lab, University Park, PA 16802, USA\\
  $^3$Harvard-Smithsonian Center for Astrophysics, MS-58, 60 Garden Street,
       Cambridge, MA 02138 USA\\
  $^4$School of Physics and Astronomy, University of St.Andrews, 
       North Haugh, St.Andrews, KY16 9SS, Scotland\\
  $^5$Institute for Astronomy, University of Vienna,
T\"{u}rkenschanzstra$\beta$e 17, A-1180 Vienna, Austria}
\date{Accepted 1998 ... . Received 1998 ... }
\pagerange{\pageref{firstpage}--\pageref{lastpage}}
\pubyear{1998}

\def\LaTeX{L\kern-.36em\raise.3ex\hbox{a}\kern-.15em
    T\kern-.1667em\lower.7ex\hbox{E}\kern-.125emX}

\begin{document}

\label{firstpage}

\maketitle

\begin{abstract}
We present high resolution optical echelle spectra and IUE 
observations during a strong flare on 1993 December 22 in the very
active, young, rapidly rotating, single K2 dwarf LQ Hya.
The initial impulsive phase of the flare, which started sometime 
between 2:42 UT and 4:07 UT, 
was characterized by strong optical continuum enhancement and blue-shifted
emission lines with broad wings.  The optical chromospheric lines
reached their maximum intensity at $\approx$5:31 UT, by which time
the blue-shift vanished and the optical continuum enhancement had 
sharply decreased.  Thereafter, the line
emission slowly decreased and the lines red-shift 
in a gradual phase that lasted at least two more hours. 
The Mg~{\sc ii} lines behaved similarly. 
Quiescent C~{\sc iv} flux levels were not recovered until 21 hours later, 
though a data gap and a possible second flare make the interpretation
uncertain. In addition to the typically flare-enhanced emission lines 
(e.g., H$\alpha$ and H$\beta$), we
observe He~{\sc i} D$_{3}$ going into emission, plus excess emission
(after subtraction of the quiescent spectrum) 
in other He~{\sc i} and several strong neutral
metal lines (e.g., Mg~{\sc i} b). 
Flare enhancement of the far UV continuum generally agrees with a 
Si~{\sc i} recombination model.  
We estimate the total flare energy, and discuss 
the broad components, asymmetries, and Doppler shifts 
seen in some of the emission lines. 

\end{abstract}


\section{Introduction}

LQ Hya (HD 82558) is a rapidly rotating 
({\it v} sin{\it i} = 25 km s$^{-1}$), single K2 dwarf, classified as a 
BY Dra variable (Fekel et al.~1986a; Fekel, Moffett \& Henry 1986b;  
Strassmeier \& Hall 1988),
with a photometric rotational period of 1.600881 days (Strassmeier et al. 1997).
Its high lithium abundance (Fekel et al.~1986a)
suggests LQ Hya has age $t < 7.5\times 10^{7}$ years (at least as young 
as the youngest Pleiades star); Vilhu, Gustafsson \& Walter (1991) even 
suggest that it may be a pre-main sequence object.
Saar, Piskunov \& Tuominen (1992, 1994), Strassmeier et al.~(1993)  
and Rice \& Strassmeier (1998)
found variable spot distribution in 
this star using Doppler-imaging, and 
widespread magnetic fields have been detected by
Saar et al.~(1992, 1994), Basri \& Marcy (1994), Donati et al.~(1997), and Donati (1998). 

LQ Hya is also (not surprisingly) a very active star, as indicated by emission 
in several chromospheric lines, including 
Ca~{\sc ii} H \& K (Fekel et al.~1986a,b; Strassmeier et al.~1990), 
Ca~{\sc ii} $\lambda$8542 (Basri \& Marcy 1994), and 
H$\alpha$ (Vilhu et al.~1991), which can also appear as 
partly (Strassmeier et al.~1993)  or completely filled-in absorption
(Fekel et al.~1986a). A filled-in He~{\sc i} D$_{3}$
line is reported by Vilhu et al.~(1991) and Saar et al.~(1997). 
Strong UV chromospheric and transition region emission lines have also been 
found by Simon \& Fekel (1987), and the star has 
been detected by ROSAT (Pye et al.~1995) and EUVE (Bowyer et al.~1996).

Flares are believed to result from the release of magnetic free energy 
stored in the corona through reconnection 
(see reviews by Mirzoyan 1984; Haisch, Strong \& Rodon\`{o} 1991).
Many types of cool stars flare (Pettersen 1989), sometimes at levels 
several orders of magnitude more energetic than
their solar counterparts.
In the dMe stars (or UV Cet type stars) optical flares are a common
phenomenon, however, in more luminous stars flares are usually only detected 
through UV or X-ray observations (e.g., Landini et al.~1986; 
H\"{u}nsch \& Reimers 1995; Ayres et al.~1994); optical 
flares are rare (Catalano 1990; 
Saar, N\"ordstrom \& Andersen 1990; Henry \& Newsom 1996).

Ambruster \& Fekel (1990) detected a strong  ultraviolet  flare on LQ Hya  in 
four continuum bands between 1250  and  1850~\AA,  while  no 
enhancement was evident in any of the chromospheric lines.
Recently, HST GHRS observations by Saar \& Bookbinder (1998) showed that
many low-level flares are present in the transition region lines of this 
star.  But consistent with its K2 spectral type and correspondingly
increased optical continuum, strong optical flares are rare on LQ Hya
(Montes et al.~1998b); Henry \& Newsom (1996), for example, saw none. 

In this paper, we report one of these rare events: 
the detection of an unusually strong optical flare in 
LQ Hya through simultaneous observations of several optical chromospheric
activity indicators:
H$\alpha$,  H$\beta$,
Na~{\sc i} D$_{1}$, D$_{2}$, He~{\sc i} D$_{3}$, Mg~{\sc i} b triplet
lines, and several UV  chromospheric and transition region lines.

In Sect. 2 we give the details of our observations and data reduction.
In Sect. 3 we describe the different aspects  of the flare
deduced from our echelle spectra, such as 
the continuum enhancement, the response of chromospheric lines 
to the flare, the variation of other photospheric lines, 
the energy released of various emission features as a function of time 
during the flare, and the line asymmetries.
Finally,  Sect. 4 gives the conclusions.

\begin{table}
\caption[]{Observing log WHT/UES (1993 December 22)
\label{tab:obslogues93}}
\begin{flushleft}
\scriptsize
\begin{tabular}{ccccl}
\noalign{\smallskip}
\hline
\noalign{\smallskip}
 UT     & JD           & {$\varphi$} &  
 S/N & Description \\
\noalign{\smallskip}
(h:m:s) & (2449343.0+) &                   &                  
     &             \\
\noalign{\smallskip}
\hline
\noalign{\smallskip}
01:14:26 & 0.55 & 0.444 &  52 & Quiescent \\
02:41:52 & 0.61 & 0.482 & 108 & Quiescent \\
04:35:48 & 0.69 & 0.532 & 145 & Impulsive \\
05:31:18 & 0.73 & 0.557 & 118 & Flare maximum \\
06:00:43 & 0.75 & 0.569 & 125 & Gradual \\
06:07:52 & 0.76 & 0.573 & 134 & " \\
06:14:20 & 0.76 & 0.576 & 132 & " \\
06:21:04 & 0.76 & 0.578 & 129 & " \\
06:29:06 & 0.77 & 0.582 & 123 & " \\
06:35:06 & 0.77 & 0.585 & 124 & " \\
06:41:03 & 0.78 & 0.587 & 129 & " \\
06:47:01 & 0.78 & 0.590 & 130 & " \\
06:53:01 & 0.79 & 0.592 & 122 & " \\
06:59:08 & 0.79 & 0.595 & 123 & " \\
07:05:06 & 0.80 & 0.598 & 122 & " \\
07:11:04 & 0.80 & 0.600 & 128 & " \\
07:17:08 & 0.80 & 0.603 & 128 & " \\
07:23:07 & 0.81 & 0.605 & 124 & " \\
07:29:05 & 0.81 & 0.608 & 129 & " \\
\noalign{\smallskip}
\hline
\end{tabular}

\end{flushleft}
\end{table}

\begin{table}
\caption[]{Continuum variation$^*$ during the flare
\label{tab:continuum}}
\begin{flushleft}
\begin{tabular}{lcccc}
\hline
\noalign{\smallskip}
 {Obs.} & $\lambda$4866 &  $\lambda$5175 
& $\lambda$5868 &  $\lambda$ 6540 \\
 (UT)   & (\%)   & (\%) & (\%) & (\%)  \\ 
\hline
\noalign{\smallskip}
%
02:42$^1$  & -   & -     & -     & -     \\
04:36             & 34  & 32    & 26    & 23    \\
05:31$^2$   & 10  & 8     & 7     & 8    \\
06:01             & 5   & 5     & 4     & 6     \\
06:29             & 6   & 7     & 5     & 4     \\
07:29             & 9   & 9     & 7     & 2     \\
\noalign{\smallskip}
\hline
\noalign{\smallskip}
\multicolumn{5}{l}{$^*$ near H$\beta$, Mg {\sc i} b, He {\sc i} D$_{3}$,
and H$\alpha$, respectively} \\
\multicolumn{5}{l}{$^1$Quiescent spectrum; $^2$Flare maximum } 
\end{tabular}
\end{flushleft}
\end{table}

\section{Observations and Data Reduction}

Echelle spectroscopic observations 
of LQ Hya were obtained 
with the 4.2m William Herschel Telescope (WHT) and the
Utrecht Echelle Spectrograph (UES) on 1993 December 22,
covering several optical chromospheric activity indicators. 
These WHT/UES spectra were obtained with echelle 31
(31.6 grooves per mm) and a 1024~x~1024 pixel TEK2 CCD as detector.
The central wavelength is 5808~\AA$\ $ covering a wavelength range
from 4842 to 7725 \AA$\ $ in a total of 44 echelle orders.
The reciprocal  dispersion ranged from 0.048 to 0.076~\AA/pixel.
In Table~\ref{tab:obslogues93} we give the observing log.
For each echelle spectrum we list
the universal time (UT), the Julian Date (JD),
the rotational phase ($\varphi$), 
and signal to noise ratio (S/N) obtained in the H$\alpha$ line region.
The rotational phase ($\varphi$) was calculated with the
ephemeris recently given by  Strassmeier et al.~(1997)
(T$_{0}$~=~2445275.0, P$_{\rm phtm}$~=~1.600881),
for alternative period determinations see also
Strassmeier et al.~(1993) and Jetsu (1993).

The spectra have been extracted using the standard IRAF 
reduction procedures 
\footnote{IRAF is distributed by the National Optical Observatory,
which is operated by the Association of Universities for Research in
Astronomy, Inc., under contract with the NSF.}
 (bias subtraction,
flat-field division, and optimal extraction of the spectra).
The wavelength calibration was obtained using  
spectra of Th-Ar lamp. 
Finally, the spectra were normalized by
a polynomial fit  to the observed continuum.


Frequent IUE SWP and LWP spectra of LQ Hya were taken between 15 and 24
Dec 1993.  The data were reduced with standard IUE software and fluxes
above background determined by simple integration (Table \ref{tab:uv_fluxes}). 
Here we analyze the NEWSIPS calibrated data, resulting 
in some small changes in the measured  
fluxes relative to our initial results (Montes et al.~1998b).


\section{Description of the Flare}


We detected a strong flare during the echelle observations of LQ Hya  
on 1993 December 22.
The temporal evolution of the flare consists of an initial 
impulsive phase which started between 2:42 UT (last 
quiescent optical spectrum) 
and  4:07 UT (end of the first IUE exposure with enhanced emission).
By the next optical spectrum (4:36 UT) 
strong increases in the chromospheric lines and continuum are seen.
The optical chromospheric lines reached maximum intensity at 5:31 UT,
by which time the continuum enhancement had already strongly decreased.
After this, the lines slowly decreased in a gradual phase that 
lasted at least until the end of the observation (07:29 UT),
i.e., $>$ 2 hours.

In the following we describe in detail the various aspects of the flare
deduced from our spectra, first exploring the time variation of the 
continuum, and then the response of the lines.
Line emission is seen both in 
the ``typical" chromospheric diagnostics (H~{\sc i} Balmer, and
He~{\sc i} D$_{3}$)  
and, after the subtraction of the quiescent spectrum, 
also in other He~{\sc i}  lines 
($\lambda$4921.9, 5015.7, and 6678.2~\AA)
and other strong lines such as the Na~{\sc i} D$_{1}$ and D$_{2}$, 
the Mg~{\sc i} b triplet and several Fe~{\sc i} and Fe~{\sc ii} lines.
Finally, we calculate the energy release during the flare 
and we analyse the broad component and the asymmetry exhibited by
some of the emission lines.


\subsection{The variation of the continuum}

Our echelle spectra show a change in depth of all the
photospheric lines due to continuum enhancement during the flare.
We have determined the contribution of the flare to the 
total observed continuum by calculating what fraction of a 
linear continuum must be added to 
the quiescent spectrum in order to reproduce the corresponding flare spectrum.
In Table \ref{tab:continuum} we give this contribution in 
representative spectral orders for several spectra
during the flare.
The maximum contribution of the flare to the continuum is reached at the
beginning of the event (the impulsive phase) and decreases thereafter. It
clearly depends on wavelength:
in the H$\beta$ line region the maximum contribution is 34 \%, 
in the He~{\sc i} D$_{3}$ line region is
26 \%, and in the  H$\alpha$ line region is 23 \%, giving a power law
index $F_{\rm cont} \propto \lambda^{-1.35}$ and an approximate blackbody
temperature of $T \approx$ 7500 K.

The continuum behaviour is thus in agreement with photometry 
of other stellar flares,  showing that the flare is initially dominated 
by strong continuum radiation, strongest at short wavelengths. 
The temperature indicated suggests that the plasma has already
cooled somewhat and thus our
``impulsive" spectrum (4:36 UT) may come somewhat late in the impulsive phase.

\subsection{The response of the optical Chromospheric lines to the flare}

To analyse the behavior 
of the various optical chromospheric lines during the flare,
we first subtracted the quiescent spectrum (UT:02:42),
as is normally done in the analysis of flare stars.
However, since the star is active,  this procedure
underestimates the total 
chromospheric contribution in these features, and ignores any
variation (due to e.g., rotational modulation or evolution) of the ``quiescent"
state. To obtain total chromospheric contribution, 
we applied the spectral subtraction technique 
(i.e., subtraction of the rotationally broadened, radial-velocity
shifted spectrum of a inactive star
chosen to match the spectral type and luminosity class of LQ Hya; 
see Montes et al.~1995a,~b,~c).
We used HD 10476 (K1V) as the inactive reference star, taken from the 
spectral library of Montes \& Mart\'{\i}n (1998).

In the case of the H$\alpha$ and H$\beta$ lines 
we have computed, for all the flare spectra, both 
the observed - quiescent (O-Q) and observed - reference (O-R) profiles 
(see Figures~\ref{fig:uesha} and ~\ref{fig:ueshb}).
For the  rest of the lines, not affected by chromospheric activity in the
quiescent spectrum, 
we studied only the (O-Q) spectra
(see Figures~\ref{fig:ueshed3},  \ref{fig:ueshe6678}, ~\ref{fig:uesna}, 
~\ref{fig:uesmg}).
In all cases, before the standard spectrum 
(quiescent or reference) is subtracted we take into account 
the contribution of the flare to the continuum (\S 3.1) 
in each spectral region, for each flare spectrum.

\subsubsection{The H$\alpha$ line}

\begin{figure*}
{\psfig{figure=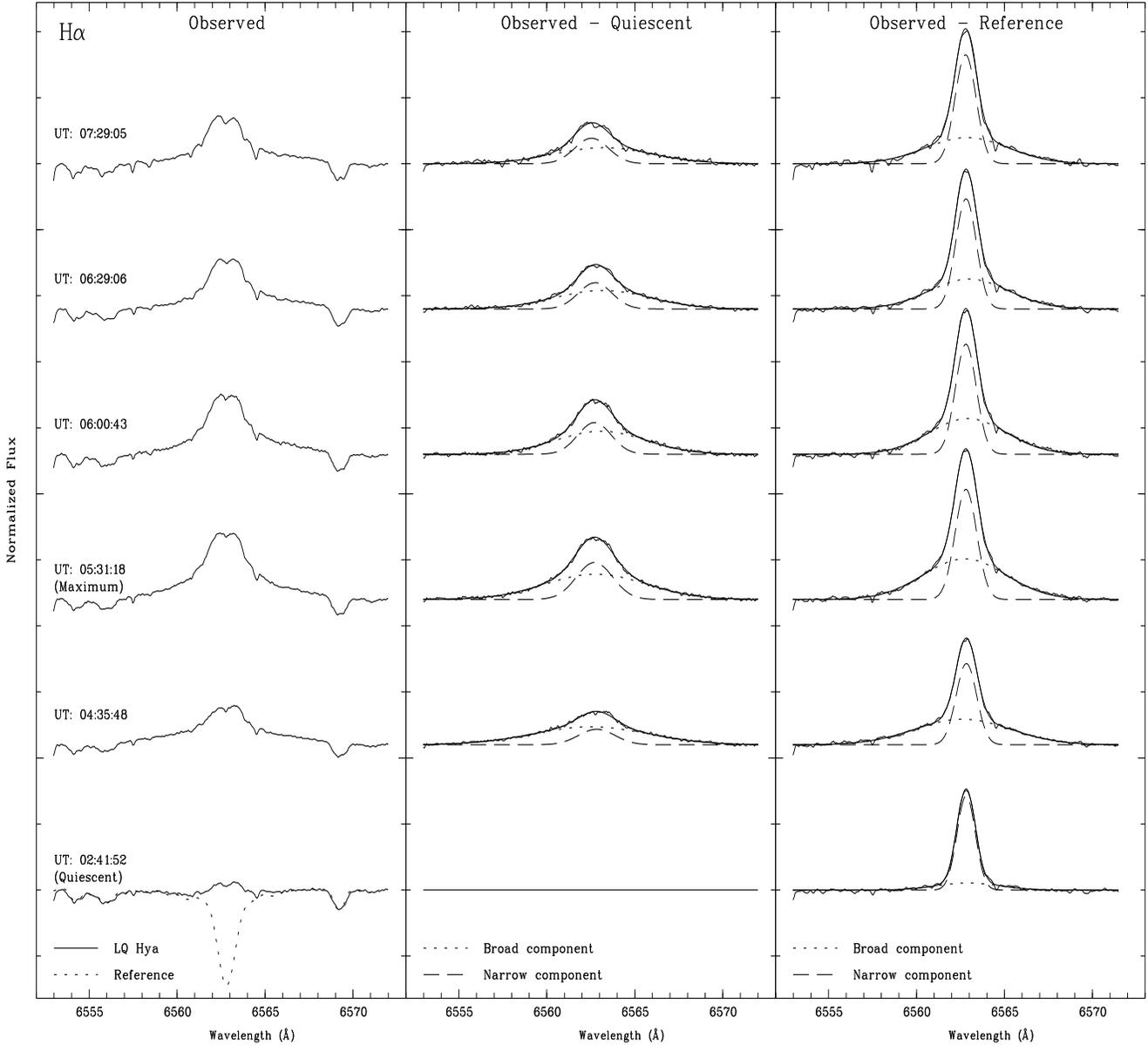,bbllx=20pt,bblly=166pt,bburx=705pt,bbury=678pt,height=16.0cm,width=17.5cm,clip=}}
\caption[ ]{H$\alpha$ observed spectra (left panel), 
after the subtraction of the quiescent spectrum (central panel) and
after the spectral subtraction (right panel)
\label{fig:uesha} }
\end{figure*}

\begin{figure*}
{\psfig{figure=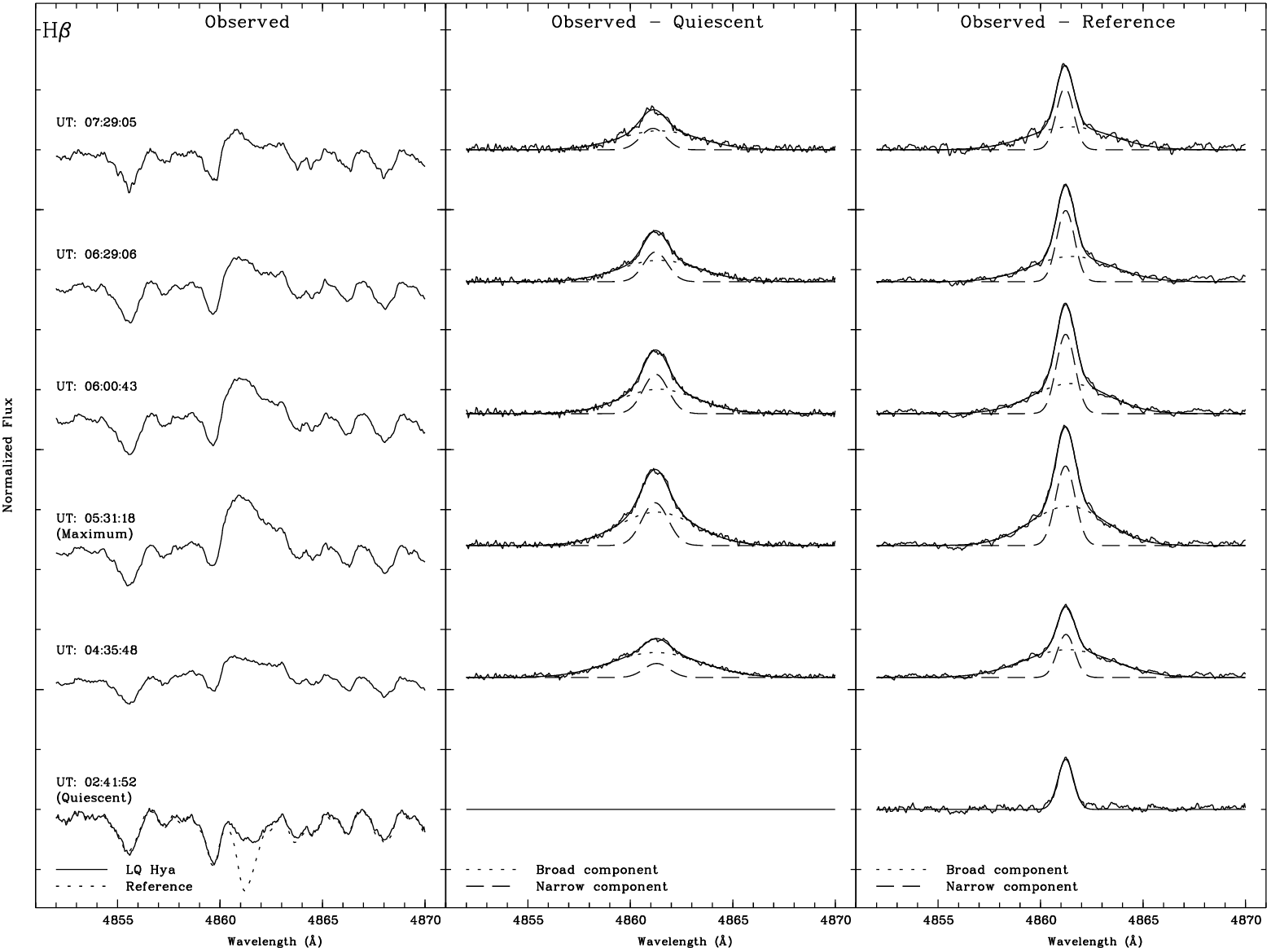,bbllx=20pt,bblly=166pt,bburx=705pt,bbury=678pt,height=16.0cm,width=17.5cm,clip=}}
\caption[ ]{H$\beta$ observed spectra (left panel), 
after the subtraction of the quiescent spectrum (central panel) and
after the spectral subtraction (right panel)
\label{fig:ueshb} }
\end{figure*}

In Fig.~\ref{fig:uesha} we have plotted the quiescent spectrum,
the reference star spectrum, and
representative spectra during the development of the flare in the
the H$\alpha$ region.
In the left panel we plot the observed spectra,
in the central panel the (O-Q) profiles,
and in the right panel the (O-R) profiles.
This figure clearly shows the conspicuous H$\alpha$ emission
enhancement, from a weak emission
above the continuum (quiescent spectrum), to a very strong
and broad emission at the maximum of the flare.
The excess H$\alpha$ emission equivalent width in the (O-R) 
spectra, increases by a factor
of $\approx$2.7 in an interval of 2.8 hours.
After the maximum the emission decreases slowly; if modeled with an 
exponential decay EW $\propto$ exp($-t/\beta$), the e-folding time 
$\beta \sim$2.5 hours in the first hour, slowing even further to 
$\beta \sim$11 hours in the second hour.

One significant aspect of these spectra is a broad emission component in the
both the (O-Q) and (O-R) profiles which
makes the total emission poorly matched by a single-Gaussian fit.
To study these profiles, we have 
therefore fitted them using
two Gaussian components: a narrow component (N) having a FWHM of
56-69 km~s$^{-1}$ and a broad component (B) with 
190 $\leq$ FWHM $\leq$ 293 km~s$^{-1}$.
In Table~\ref{tab:measuresha} we list the parameters (I, FWHM, EW)
of the broad and narrow components.
As can be seen in this table, the contribution of the B component
to the total EW and FWHM of the line is a maximum in the
impulsive phase.  The line profiles are also asymmetric,
and the two Gaussian fit is optimized when the broad component
is blue- or red-shifted with respect to the narrow component
(see the $\Delta \lambda$ = $\lambda$$_{\rm N}$ - $\lambda$$_{\rm B}$
value in Table~\ref{tab:measuresha}).  We discuss this further in \S 3.5.

In Table~\ref{tab:measuresha} we also give,
for the total subtracted spectra,
the peak emission intensity (I),
 the excess H$\alpha$ emission equivalent width (EW(H$\alpha$)), and
absolute fluxes at the stellar surface, logF$_{\rm S}$(H$\alpha$),
in erg cm$^{-2}$ s$^{-1}$ 
obtained with the calibration of Hall (1996).
The time evolution of the EW( H$\alpha$) during the flare is
shown in Fig.~\ref{fig:lqhya_ews}.

\begin{table*}
\caption[]{H$\alpha$ parameters during the flare in the
subtracted profiles for the two Gaussian component fits and 
for the total emission
\label{tab:measuresha}}
\begin{flushleft}
\begin{tabular}{lcccccccccccccc}
\hline
        &   
\multicolumn{4}{c}{H$\alpha$ broad component} &  &
\multicolumn{4}{c}{H$\alpha$ narrow component} \\
\cline{2-5}\cline{7-10}
\noalign{\smallskip}
 {Obs.} & 
{I} & {\scriptsize FWHM} & EW$_{\rm B}$ & $_{\rm B}$/$_{\rm T}$ & &
{I} & {\scriptsize FWHM} & EW$_{\rm N}$ & $_{\rm N}$/$_{\rm T}$ &
$\Delta \lambda$ & I$_{\rm T}$ & EW$_{\rm T}$ & $\log {\rm F}_{\rm T}$ \\
 (UT)   &       
    & {\scriptsize (\AA)} & {\scriptsize (\AA)} & (\%) & &
    & {\scriptsize (\AA)} & {\scriptsize (\AA)} & (\%) & 
 ($\lambda$$_{\rm N}$ - $\lambda$$_{\rm B}$) & & {\scriptsize (\AA)} & \\
\hline
{\bf Obs. - Quiescent} \\
\hline
\noalign{\smallskip}
%
02:42 (Quiescent) & -   & -     & -     & -    &
                & -     & -     & -     & -    & -     & 0.000 & 0.000 & 0.00 \\
04:36           & 0.137 & 7.709 & 1.098 & 78.4 &
                & 0.117 & 2.414 & 0.304 & 45.2 & +0.34 & 0.254 & 1.401 & 6.70 \\
05:31 (Maximum) & 0.192 & 6.766 & 1.371 & 65.4 &
                & 0.281 & 2.427 & 0.725 & 34.6 & -0.06 & 0.471 & 2.096 & 6.88 \\
06:01           & 0.174 & 6.021 & 1.113 & 66.1 &
                & 0.240 & 2.227 & 0.570 & 33.9 & -0.36 & 0.417 & 1.683 & 6.78 \\
06:29           & 0.140 & 6.414 & 0.948 & 66.7 &
                & 0.199 & 2.238 & 0.473 & 33.3 & -0.42 & 0.331 & 1.421 & 6.71 \\
07:29           & 0.121 & 6.567 & 0.837 & 64.3 &
                & 0.193 & 2.267 & 0.465 & 35.7 & -0.66 & 0.315 & 1.302 & 6.67 \\
\noalign{\smallskip}
\hline
{\bf Obs. - Reference} \\
\hline
\noalign{\smallskip}
%
02:42 (Quiescent)&0.053 & 4.168 & 0.236 & 20.3 &
                & 0.714 & 1.221 & 0.927 & 79.7 & -0.12 & 0.753 & 1.163 & 6.62 \\
04:36           & 0.194 & 6.414 & 1.316 & 59.4 &
                & 0.615 & 1.371 & 0.898 & 40.6 & +0.21 & 0.802 & 2.214 & 6.90 \\
05:31 (Maximum) & 0.307 & 5.532 & 1.809 & 58.5 &
                & 0.835 & 1.442 & 1.281 & 41.5 & +0.07 & 1.132 & 3.090 & 7.04 \\
06:01           & 0.270 & 5.153 & 1.481 & 54.4 &
                & 0.834 & 1.397 & 1.240 & 45.6 & -0.11 & 1.085 & 2.721 & 6.99 \\
06:29           & 0.226 & 5.326 & 1.280 & 51.2 &
                & 0.835 & 1.221 & 1.221 & 48.8 & -0.17 & 1.043 & 2.501 & 6.95 \\
07:29           & 0.201 & 5.449 & 1.167 & 48.6 &
                & 0.822 & 1.397 & 1.233 & 51.4 & -0.09 & 1.006 & 2.400 & 6.93 \\
%
\noalign{\smallskip}
\hline
\noalign{\smallskip}
\end{tabular}

\end{flushleft}
\end{table*}

\begin{table*}
\caption[]{H$\beta$ parameters during the flare in the
subtracted profiles for the two Gaussian component fits and 
for the total emission
\label{tab:measureshb}}
\begin{flushleft}
\begin{tabular}{lcccccccccccccc}
\noalign{\smallskip}
\hline
        &   
\multicolumn{4}{c}{H$\beta$ broad component} &  &
\multicolumn{4}{c}{H$\beta$ narrow component} \\
\cline{2-5}\cline{7-10}
\noalign{\smallskip}
 {Obs.} & 
{I} & {\scriptsize FWHM} & EW$_{\rm B}$ & $_{\rm B}$/$_{\rm T}$ & &
{I} & {\scriptsize FWHM} & EW$_{\rm N}$ & $_{\rm N}$/$_{\rm T}$ &
$\Delta \lambda$ & I$_{\rm T}$ & EW$_{\rm T}$ & H$\alpha$/H$\beta$ \\
 (UT)   &       
    & {\scriptsize (\AA)} & {\scriptsize (\AA)} & (\%) & &
    & {\scriptsize (\AA)} & {\scriptsize (\AA)} & (\%) & 
($\lambda$$_{\rm N}$ - $\lambda$$_{\rm B}$) & & {\scriptsize (\AA)} & \\
\hline
{\bf Obs. - Quiescent} \\
\hline
\noalign{\smallskip}
%
02:42 (Quiescent) & -   & -     & -     & -    &
                & -     & -     & -     & -    & -     & 0.000 & 0.000 & 0.00 \\
04:36           & 0.208 & 5.185 & 1.142 & 86.3 &
                & 0.117 & 1.450 & 0.181 & 13.7 & +0.03 & 0.326 & 1.323 & 1.143\\
05:31 (Maximum) & 0.279 & 4.440 & 1.142 & 69.7 &
                & 0.357 & 1.507 & 0.572 & 30.3 & -0.08 & 0.644 & 1.888 & 1.498\\
06:01           & 0.203 & 4.429 & 0.957 & 65.5 &
                & 0.328 & 1.439 & 0.503 & 34.5 & -0.15 & 0.532 & 1.450 & 1.253\\
06:29           & 0.178 & 4.599 & 0.870 & 70.7 &
                & 0.248 & 1.361 & 0.359 & 29.3 & -0.20 & 0.429 & 1.230 & 1.246\\
07:29           & 0.158 & 4.569 & 0.767 & 73.9 &
                & 0.178 & 1.428 & 0.271 & 26.1 & -0.30 & 0.366 & 1.038 & 1.353\\
\noalign{\smallskip}
\hline
{\bf Obs. - Reference} \\
\hline
\noalign{\smallskip}
%
02:42 (Quiescent)&-     & -     & -     & -    &
                & -     & -     & -     & -    & -     & 0.437 & 0.398 & 3.152\\
04:36           & 0.232 & 5.102 & 1.253 & 76.5 &
                & 0.359 & 1.004 & 0.384 & 23.5 & -0.09 & 0.610 & 1.636 & 1.460\\
05:31 (Maximum) & 0.327 & 4.308 & 1.501 & 65.5 &
                & 0.663 & 1.118 & 0.789 & 34.5 & -0.13 & 1.003 & 2.290 & 1.455\\
06:01           & 0.250 & 4.343 & 1.156 & 60.4 &
                & 0.661 & 1.076 & 0.757 & 39.6 & -0.23 & 0.920 & 1.913 & 1.534\\
06:29           & 0.211 & 4.627 & 1.036 & 61.4 &
                & 0.594 & 1.028 & 0.650 & 38.6 & -0.30 & 0.814 & 1.686 & 1.600\\
07:29           & 0.191 & 4.713 & 0.955 & 63.5 &
                & 0.514 & 1.000 & 0.548 & 36.5 & -0.28 & 0.718 & 1.503 & 1.723\\
%
\noalign{\smallskip}
\hline
\noalign{\smallskip}
\end{tabular}

\end{flushleft}
\end{table*}

\subsubsection{The H$\beta$ line}

Fig.~\ref{fig:ueshb} is as Fig.~\ref{fig:uesha} for the H$\beta$ line.
In this case the line changes from mostly filled-in absorption 
(quiescent), to a strong and broad emission line at flare maximum. 
The excess H$\beta$ emission EW in (O-R) 
spectra increases by a factor of 5.8 from the quiescent level to the maximum,
a considerably larger enhancement than in the H$\alpha$ line.
However, during the gradual phase the emission 
declines slightly more rapidly than the H$\alpha$ emission 
(see Fig. \ref{fig:lqhya_ews}), with $\beta = 2.3$ hours in the first hour of
decay, slowing to $\beta = 5.9$ hours in the second. The more rapid decay
of H$\beta$ has also been observed in other solar and stellar flares
(Johns-Krull et al.~1997). 

Like H$\alpha$,  very broad wings are visible in the
(O-R) H$\beta$ profiles, and the results of double Gaussian fits are 
given in Table~\ref{tab:measureshb}.
The contribution and FWHM of the broad component again reach a
maximum in the impulsive phase and 
the profiles show an asymmetry similar to that seen in H$\alpha$. 

\subsubsection{The H$\alpha$/H$\beta$ Balmer decrement  }

The Balmer decrement 
can be an important indication of physical parameters
such as electron density (Kunkel 1970).
It is commonly tabulated for stellar flares with respect
to the H$\gamma$ line (Hawley \& Pettersen 1991).  
Lacking H$\gamma$ data in our spectra, 
we have instead calculated the Balmer decrement 
from the EW(H$\alpha$)/EW(H$\beta$) ratio, 
assuming that the LQ Hya continua at H$\alpha$ and H$\beta$ have a ratio
appropriate to a blackbody at T$_{eff}$ = 4900 K 
(F$_{\lambda 6563}$/F$_{\lambda 4861}$ = 1.08). The values obtained
are given in Table~\ref{tab:measureshb}. 

We found that the H$\alpha$/H$\beta$ Balmer decrement 
is shallower during the flare, changing from 3.15 at the quiescent state
to 1.46 at the impulsive and maximum  phases of the flare, 
indicating (unsurprisingly) a significant change in the properties of hydrogen 
emitting regions during the flare.
In all the flares on dMe stars tabulated by Hawley \& Pettersen (1991), 
the decrement for H$\beta$ and higher Balmer lines
was also shallower during the flare.
The H$\alpha$/H$\beta$ ratio can fall
to $<1$ during large stellar flares (Kunkel 1970) and in
the largest solar flares (Zirin \& Ferland 1980).
However, in other stellar flares steeper decrements (2.0) 
are reported (Worden et al.~1984) and in the solar flares observed
by Johns-Krull et al.~(1997) this ratio only changed from 2.5 to 1.45.

\subsubsection{The He~{\sc i} D$_{3}$ and other He~{\sc i} lines}

\begin{figure}
{\psfig{figure=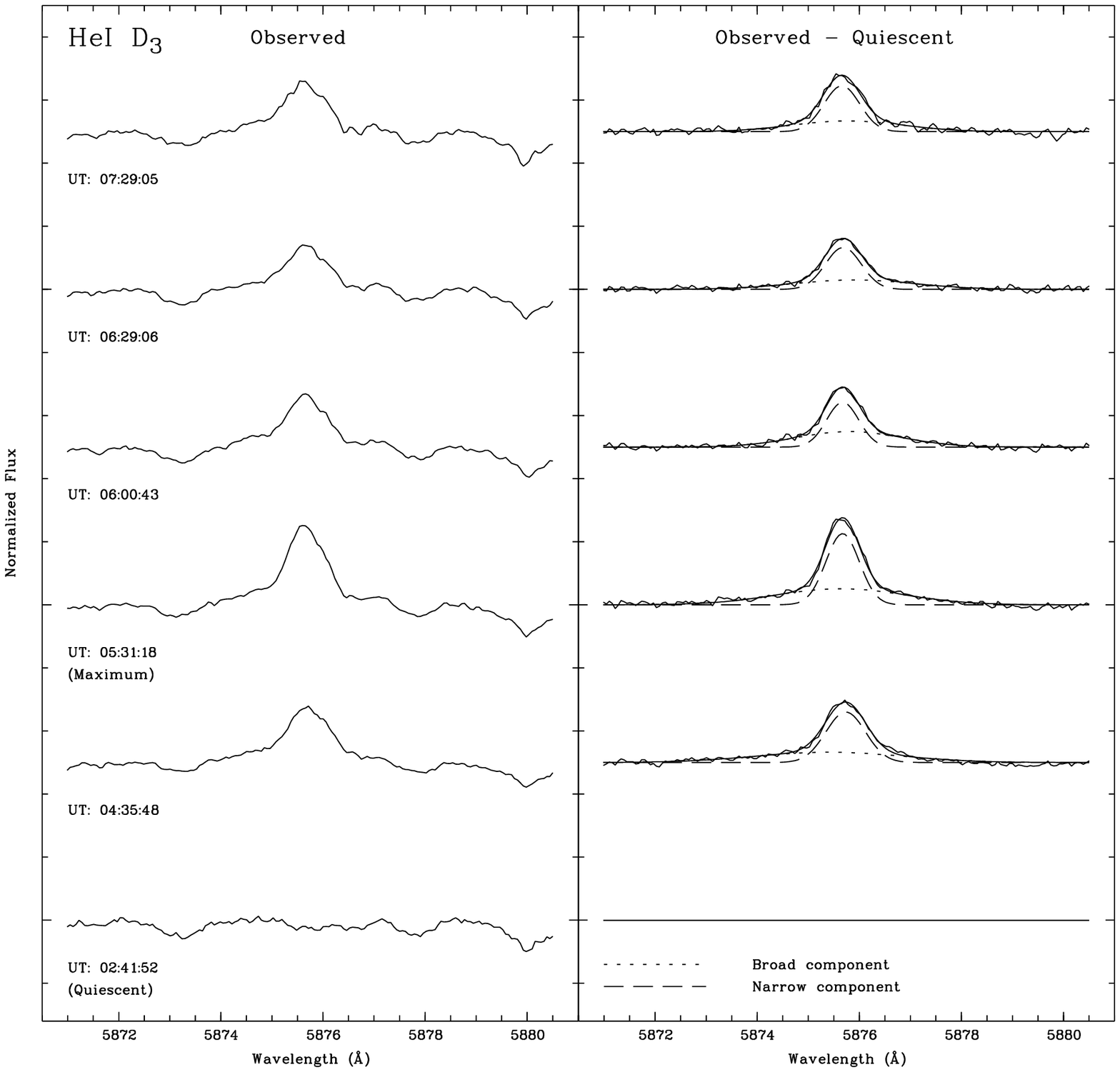,bbllx=40pt,bblly=166pt,bburx=578pt,bbury=680pt,height=12.0cm,width=8.4cm,clip=}}
\caption[ ]{He~{\sc i} D$_{3}$ observed spectra (left panel) and  
after the subtraction of the quiescent spectrum (right panel) 
\label{fig:ueshed3} }
\end{figure}

In Fig.~\ref{fig:ueshed3} we show the quiescent spectrum and 
representative flare spectra in the 
the He~{\sc i} D$_{3}$ region; the 
left panel displays the observed spectra and 
in the right panel, the (O-Q) profiles are plotted.
The He~{\sc i} D$_{3}$ line, completely filled-in in quiescence, goes 
into emission during the flare,
reaching a maximum intensity (0.346) at the same time that the Balmer lines. 
The subtracted profiles again show broad wings; an analysis 
like that used with H$\alpha$ and H$\beta$ yields
the parameters in Table~\ref{tab:measuresheid3}.

This line is a well known diagnostic of flares in the Sun and in other stars.
In the Sun the  D$_{3}$ line appears as absorption 
in plage and weak flares and as emission in strong flares (Zirin 1988).
The He~{\sc i} D$_{3}$ feature is typically in emission in dMe stars
(e.g., Pettersen, Evans \& Coleman 1984);  
in flare like events in UV Ceti flare stars
(Kunkel 1970; Bopp \& Moffett 1973);
in strong flares in more luminous, active stars such as
the RS CVn systems II Peg 
(Huenemoerder \& Ramsey 1987; Montes et al.~1997; 
Berdyugina, Ilyin \& Tuominen 1998)
and UX Ari (Montes et al.~1996b; 1997),
the weak-lined T Tauri star V410 Tau (Welty \& Ramsey 1998),  and in 
the active G5V $\kappa$ Cet (Robinson \& Bopp 1987).
Thus, the detection of prominent D$_{3}$ 
emission indicates that we are observing a very strong flare in LQ Hya.

\begin{figure}
{\psfig{figure=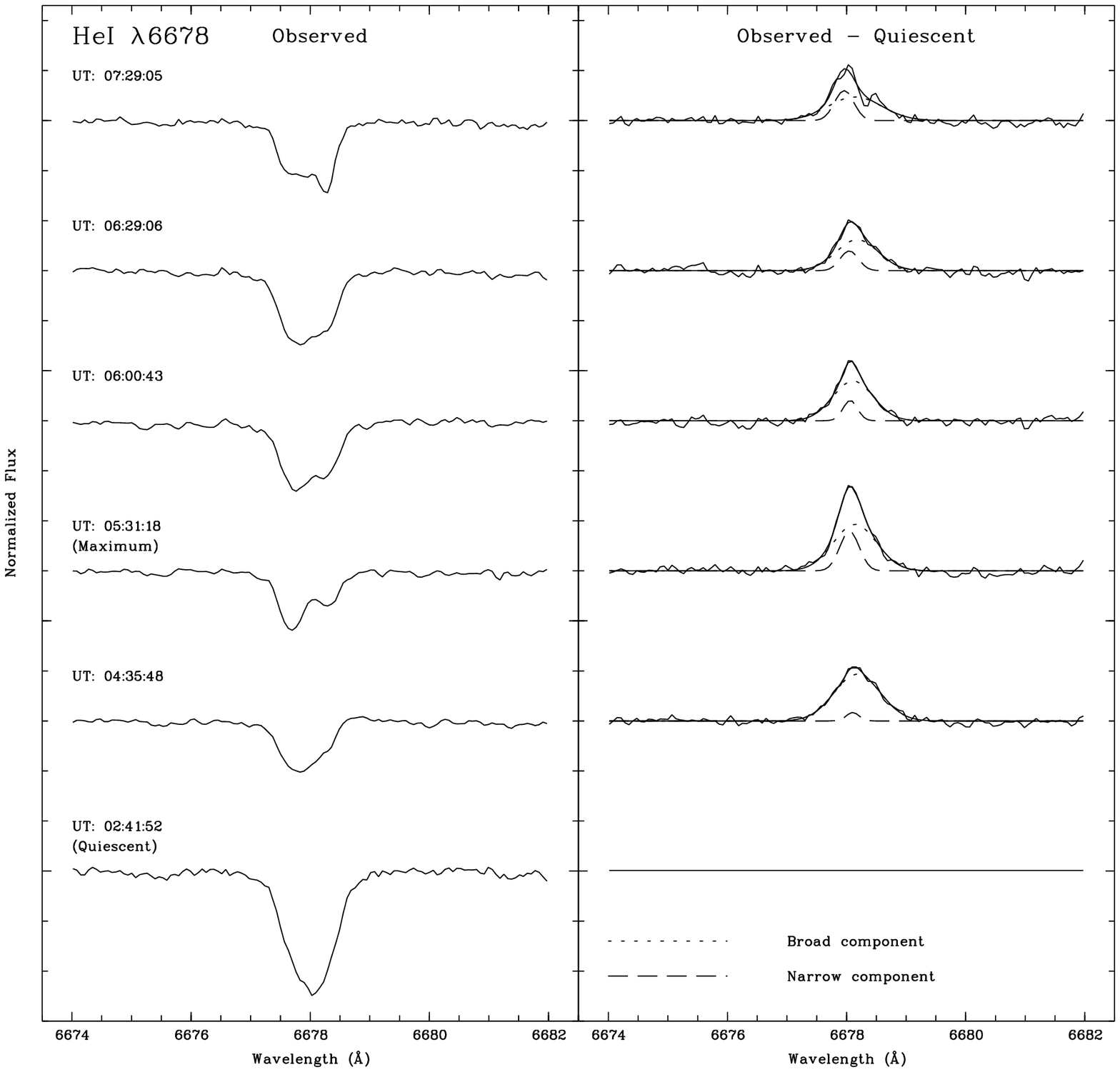,bbllx=40pt,bblly=166pt,bburx=578pt,bbury=680pt,height=12.0cm,width=8.4cm,clip=}}
\caption[ ]{He~{\sc i} $\lambda$6678 observed spectra (left panel) and  
after the subtraction of the quiescent spectrum (right panel) 
\label{fig:ueshe6678} }
\end{figure}

Other He~{\sc i} lines have been also reported in stellar flares 
(Bopp \& Moffett 1973; Hawley \& Pettersen (1991);
Abdul-Aziz et al. 1995; Abranin et al. 1998).
In our spectra, after the subtraction of the quiescent spectra, 
we have also found an excess emission in other He~{\sc i} lines
at 4921.93, 5015.68 and 6678.15~\AA. 
In particular, He {\sc i} $\lambda$6678.15 appears superimposed on the 
Fe {\sc i} $\lambda$6677.99 absorption line 
and the excess emission can be seen
even in the observed spectra (see Fig.~\ref{fig:ueshe6678} left panel).
When the quiescent spectra is subtracted, excess He~{\sc i} emission 
is clearly seen (see Fig.~\ref{fig:ueshe6678} right panel),
with a maximum peak intensity of 0.17.
The profiles have been again fitted with two Gaussians; 
the corresponding parameters are given in Table~\ref{tab:measureshei6678}.
The temporal evolution of this He~{\sc i} line during the flare 
is similar to the D$_{3}$ line.

\begin{figure}
{\psfig{figure=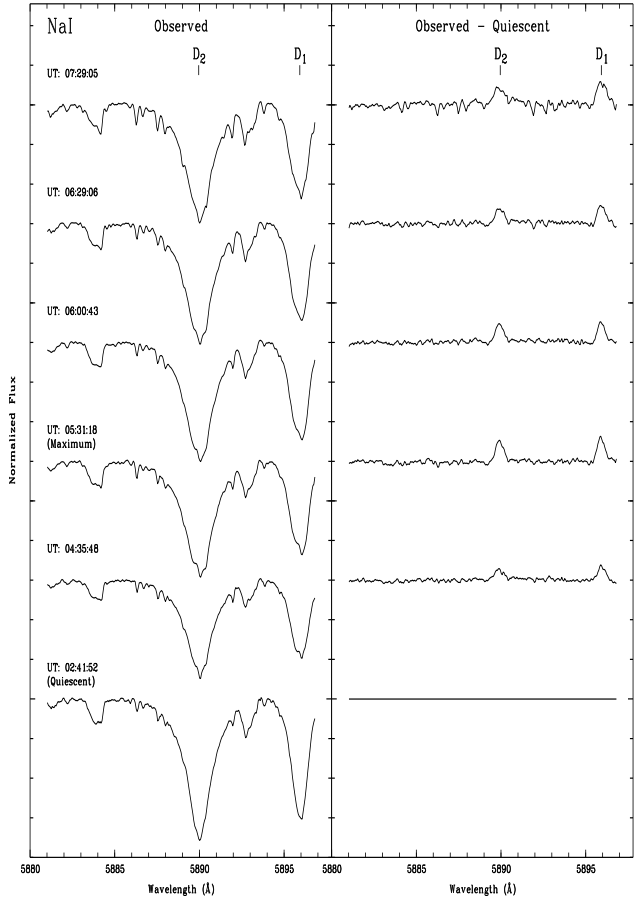,bbllx=40pt,bblly=166pt,bburx=578pt,bbury=680pt,height=12.0cm,width=8.4cm,clip=}}
\caption[ ]{Spectra near the Na~{\sc i} D$_{1}$ and D$_{2}$ lines before 
(left panel) and  
after subtracting the quiescent spectrum (right panel) 
\label{fig:uesna} }
\end{figure}
\begin{figure}
{\psfig{figure=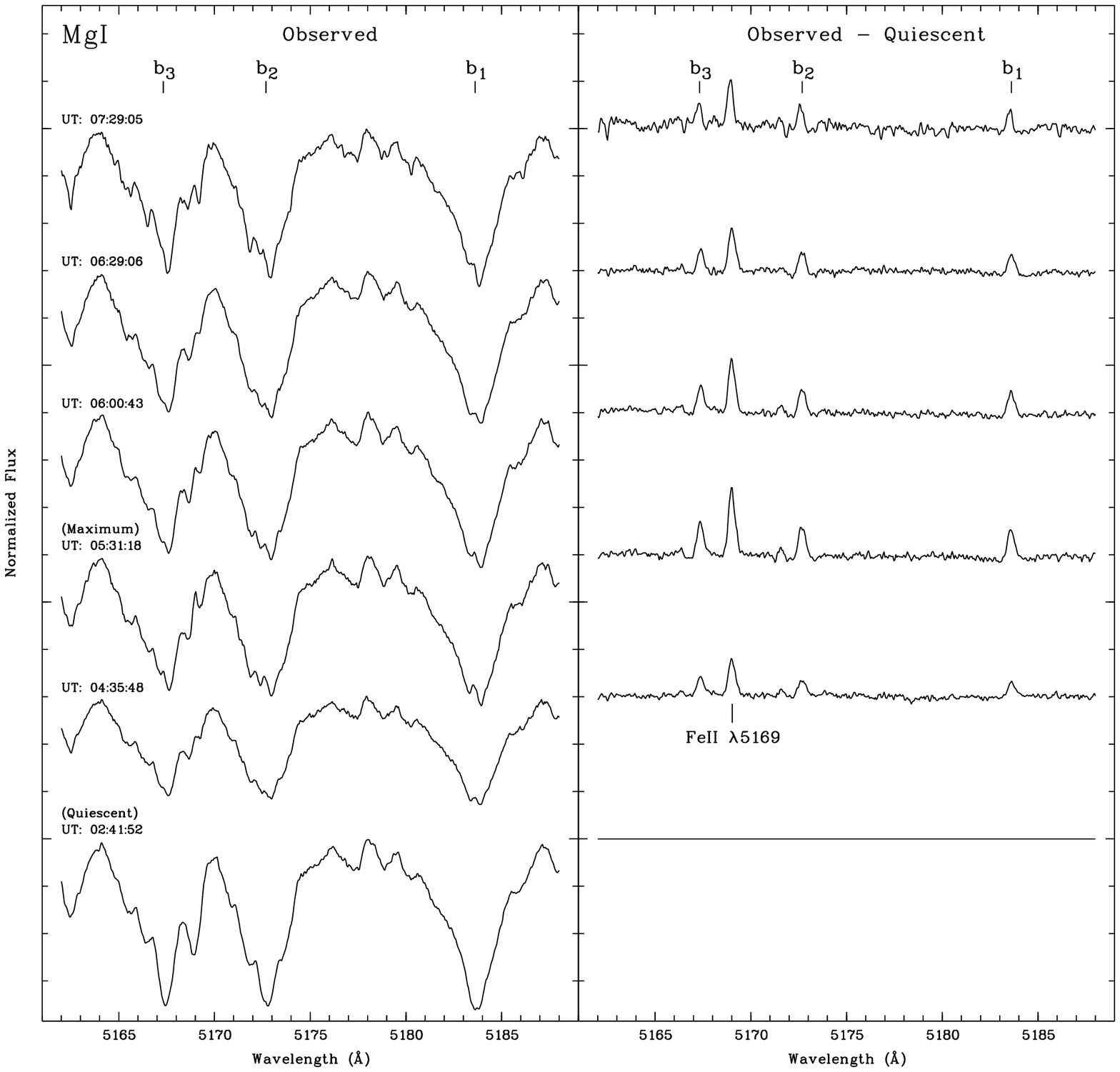,bbllx=40pt,bblly=166pt,bburx=578pt,bbury=680pt,height=12.0cm,width=8.4cm,clip=}}
\caption[ ]{Spectra near the Mg~{\sc i} b triplet lines before 
(left panel) and  
after the subtraction of the quiescent spectrum (right panel) 
\label{fig:uesmg} }
\end{figure}

\subsubsection{The Na~{\sc i} D$_{1}$ and D$_{2}$ lines}

The Na~{\sc i} D$_{1}$ and D$_{2}$ resonance lines at
5895.9~\AA\ and 5890.0~\AA\ are collisionally-controlled in the
atmospheres of late-type stars and thus provide information about
chromospheric activity (see Montes et al.~1996b, 1997, 1998a).  Recent
models of these lines for M dwarfs have been made 
by Andretta, Doyle, \& Byrne (1997).
Using model solar flares, Falchi, Falciani \& Smaldone (1990) found that the
Na~{\sc i} D$_{2}$ line shows modifications only in the core of its profile.

Subtraction of the modified reference star from the quiescent 
spectrum reveals that excess emission
in the Na~{\sc i} D$_{1}$ and D$_{2}$ lines is very small; 
we have therefore analysed only the (O-Q) spectra to study the
flare in these lines.
As can be seen in Fig.~\ref{fig:uesna},  the lines show a clear 
filling-in of the core its profile, reaching its maximum intensity (0.13)
at the same time as the other chromospheric lines.
The peak intensity (I), FWHM, and EW for both lines are given in 
Table~\ref{tab:measuresna}.

\subsubsection{The Mg~{\sc i} b triplet lines}

The strong Mg~{\sc i} b triplet  $\lambda$$\lambda$5167, 5172, 5183
is formed in the lower chromosphere and the
region of temperature minimum and they are good diagnostics of
activity (Basri, Wilcots \& Stout 1989; 
Gunn \& Doyle 1997; Gunn, Doyle \& Houdebine 1997).
In some stellar flares this lines exhibit a central reversal 
(Bopp \& Moffett 1973; Mochnacki \& Schommer 1979;
Abdul-Aziz et al. 1995; Abranin et al. (1998).
In the quiescent spectrum very strong absorption lines are observed, without 
evidence of filling-in by activity. In the 
flare spectra, however, a small reversal is observed 
in line cores (Fig.~\ref{fig:uesmg} left panel).
After the subtraction of the quiescent spectra, excess emission is clearly 
visible, both in the three Mg~{\sc i} b lines (Fig.~\ref{fig:uesmg} right 
panel), and (even more intensely) in the Fe~{\sc ii} feature at 5169.0~\AA.
The measured parameters for these lines are given in 
Table~\ref{tab:measuresmg}.



\subsubsection{The variation of other photospheric lines}

The upper photosphere of the late-type stars is also affected by the
chromospheric activity, as seen in the filling-in of 
strong photospheric lines
with relatively low excitation potentials in active dwarfs 
(Basri et al.~1989) and pre-main sequence stars (Finkenzeller \& Basri 1987).
During moderately strong solar flares a large number of
photospheric lines are filled-in when the quiescent spectrum is subtracted
(Acampa et al.~1982; Mauas 1990; Johns-Krull et al.~1997).
Filling-in of some metal lines are also reported in earlier 
observations of flares
of the UV Cet-type stars by Mochnacki \& Schommer (1979; 
and references therein), Hawley \& Pettersen (1991),
and in recently observations by
Abdul-Aziz et al. (1995) and Abranin et al. (1998).   

We have observed a slight filling in of many
photospheric absorption lines in our spectra during the flare.
These include all the lines reported to be activity-sensitive 
by Basri et al.~(1989) and Finkenzeller \& Basri (1987), 
 many of the lines appearing in 
the high resolution spectra of a moderate solar flare (Johns-Krull et al.~1997),
 all the lines identified in the spectra of a more energetic 
solar flare (Acampa et al.~1982),
and some of the lines reported in stellar flares 
(Mochnacki \& Schommer 1979;
Abdul-Aziz et al. 1995; Abranin et al. (1998).
The lines with the largest filling in are those of multiplet 42 of 
Fe {\sc ii} (4943.9, 5018.4, 5169.0~\AA).
Other lines with significant filling are the lines of 
multiplet 48  (5316.8, 5362.9~\AA)  and 
49 (5197.6, 5234.6, 5276.0, 5316.6~\AA) of Fe {\sc ii}.

\begin{table*}
\caption[]{He~{\sc i} D$_{3}$ parameters during the flare in the
(Observed - Quiescent) profiles for the two Gaussian component 
fits and for the total emission
\label{tab:measuresheid3}}
\begin{flushleft}
\begin{tabular}{lcccccccccccccc}
\hline
        &   
\multicolumn{4}{c}{He~{\sc i} D$_{3}$ broad component} &  &
\multicolumn{4}{c}{He~{\sc i} D$_{3}$ narrow component} \\
\cline{2-5}\cline{7-10}
\noalign{\smallskip}
 {Obs.} & 
{I} & {\scriptsize FWHM} & EW$_{\rm B}$ & $_{\rm B}$/$_{\rm T}$ & &
{I} & {\scriptsize FWHM} & EW$_{\rm N}$ & $_{\rm N}$/$_{\rm T}$ &
$\Delta \lambda$ & I$_{\rm T}$ & EW$_{\rm T}$  \\
 (UT)   &       
    & {\scriptsize (\AA)} & {\scriptsize (\AA)} & (\%) & &
    & {\scriptsize (\AA)} & {\scriptsize (\AA)} & (\%) & 
($\lambda$$_{\rm N}$ - $\lambda$$_{\rm B}$) & & {\scriptsize (\AA)}  \\
\hline
\noalign{\smallskip}
%
02:42 (Quiescent) & -   & -     & -     & -    &
                & -     & -     & -     & -    & -     & 0.000 & 0.000  \\
04:36           & 0.033 & 3.400 & 0.117 & 43.0 &
                & 0.160 & 0.912 & 0.155 & 57.0 & +0.32 & 0.198 & 0.272   \\
05:31 (Maximum) & 0.051 & 2.981 & 0.161 & 46.5 &
                & 0.226 & 0.772 & 0.185 & 53.5 & +0.03 & 0.272 & 0.346   \\
06:01           & 0.049 & 2.450 & 0.128 & 53.3 &
                & 0.142 & 0.745 & 0.112 & 46.7 & -0.13 & 0.190 & 0.241   \\
06:29           & 0.030 & 2.802 & 0.088 & 43.1 &
                & 0.133 & 0.825 & 0.117 & 56.9 & -0.21 & 0.161 & 0.205   \\
07:29           & 0.034 & 2.554 & 0.093 & 42.0 &
                & 0.145 & 0.826 & 0.128 & 58.0 & -0.11 & 0.184 & 0.220   \\
\noalign{\smallskip}
\hline
\noalign{\smallskip}
\end{tabular}

\end{flushleft}
\end{table*}

\begin{table*}
\caption[]{He~{\sc i} $\lambda$6678 parameters during the flare in the
(O - Q) profiles for the two Gaussian component 
fits and for the total emission
\label{tab:measureshei6678}}
\begin{flushleft}
\begin{tabular}{lcccccccccccccc}
\hline
        &   
\multicolumn{4}{c}{He~{\sc i} $\lambda$6678 broad component} &  &
\multicolumn{4}{c}{He~{\sc i} $\lambda$6678 narrow component} \\
\cline{2-5}\cline{7-10}
\noalign{\smallskip}
 {Obs.} & 
{I} & {\scriptsize FWHM} & EW$_{\rm B}$ & $_{\rm B}$/$_{\rm T}$ & &
{I} & {\scriptsize FWHM} & EW$_{\rm N}$ & $_{\rm N}$/$_{\rm T}$ &
$\Delta \lambda$ & I$_{\rm T}$ & EW$_{\rm T}$  \\
 (UT)   &       
    & {\scriptsize (\AA)} & {\scriptsize (\AA)} & (\%) & &
    & {\scriptsize (\AA)} & {\scriptsize (\AA)} & (\%) & 
($\lambda$$_{\rm N}$ - $\lambda$$_{\rm B}$) & & {\scriptsize (\AA)}  \\
\hline
\noalign{\smallskip}
%
02:42 (Quiescent) & -   & -     & -     & -    &
                & -     & -     & -     & -    & -     & 0.000 & 0.000  \\
04:36           & 0.094 & 0.911 & 0.091 & 95.4 &
                & 0.017 & 0.249 & 0.004 & 04.6 & -0.08 & 0.106 & 0.095   \\
05:31 (Maximum) & 0.093 & 0.857 & 0.085 & 72.4 &
                & 0.079 & 0.384 & 0.032 & 27.6 & -0.09 & 0.171 & 0.117   \\
06:01           & 0.080 & 0.797 & 0.067 & 84.5 &
                & 0.041 & 0.285 & 0.012 & 15.5 & -0.08 & 0.120 & 0.080   \\
06:29           & 0.062 & 0.833 & 0.055 & 78.3 &
                & 0.040 & 0.360 & 0.015 & 21.7 & -0.13 & 0.101 & 0.070   \\
07:29           & 0.047 & 1.060 & 0.053 & 68.6 &
                & 0.060 & 0.381 & 0.024 & 31.4 & -0.18 & 0.111 & 0.078   \\
\noalign{\smallskip}
\hline
\noalign{\smallskip}
\end{tabular}

\end{flushleft}
\end{table*}

\begin{table*}
\caption[]{Na~{\sc i} D line parameters 
during the flare in the (O - Q) profile
\label{tab:measuresna}}
\begin{flushleft}
\begin{tabular}{lccccccc}
\hline
        &   
\multicolumn{3}{c}{Na~{\sc i} D$_{2}$} &  &
\multicolumn{3}{c}{Na~{\sc i} D$_{1}$} \\
\cline{2-4}\cline{6-8}
\noalign{\smallskip}
 {Obs.} & 
{I} & {FWHM} & EW & &
{I} & {FWHM} & EW \\
 (UT)   &       
    & {\scriptsize (\AA)} & {\scriptsize (\AA)} & & 
    & {\scriptsize (\AA)} & {\scriptsize (\AA)} \\
\hline
\noalign{\smallskip}
%
02:42$^1$  &0.000& 0.000 & 0.000 &
                & 0.000 & 0.000 & 0.000 \\  
04:36           & 0.058 & 0.674 & 0.036 & 
                & 0.078 & 0.549 & 0.038 \\
05:31$^2$ & 0.106 & 0.584 & 0.059 & 
                & 0.127 & 0.487 & 0.065 \\
06:01           & 0.095 & 0.600 & 0.057 & 
                & 0.104 & 0.528 & 0.056 \\
06:29           & 0.074 & 0.718 & 0.048 & 
                & 0.092 & 0.657 & 0.052 \\
07:29           & 0.090 & 0.915 & 0.063 & 
                & 0.117 & 0.666 & 0.074 \\ 
\noalign{\smallskip}
\hline
\noalign{\smallskip}
\multicolumn{8}{l}{$^1$Quiescent spectrum; $^2$Flare maximum } 
\end{tabular}

\end{flushleft}
\end{table*}

\begin{table*}
\caption[]{Mg~{\sc i} b triplet and Fe~{\sc ii} $\lambda$5169 
line parameters during the flare in the (O - Q) profile
\label{tab:measuresmg}}
\begin{flushleft}
\begin{tabular}{lccccccccccccccccccc}
\hline
        &   
\multicolumn{3}{c}{Mg~{\sc i} b$_{3}$} &  &
\multicolumn{3}{c}{Mg~{\sc i} b$_{2}$} &  &
\multicolumn{3}{c}{Mg~{\sc i} b$_{1}$} &  & 
\multicolumn{3}{c}{Fe~{\sc ii} $\lambda$5169 } \\
\cline{2-4}\cline{6-8}\cline{10-12}\cline{14-16}
\noalign{\smallskip}
 {Obs.} & 
{I} & {FWHM} & EW & &
{I} & {FWHM} & EW & &
{I} & {FWHM} & EW & &
{I} & {FWHM} & EW   \\
 (UT)   &       
    & {\scriptsize (\AA)} & {\scriptsize (\AA)} & & 
    & {\scriptsize (\AA)} & {\scriptsize (\AA)} & & 
    & {\scriptsize (\AA)} & {\scriptsize (\AA)} & &
    & {\scriptsize (\AA)} & {\scriptsize (\AA)}  \\
\hline
\noalign{\smallskip}
%
02:42$^1$       &0.000& 0.000 & 0.000 &
                & 0.000 & 0.000 & 0.000 &
                & 0.000 & 0.000 & 0.000 &
                & 0.000 & 0.000 & 0.000  \\
04:36           & 0.078 & 0.531 & 0.044 & 
                & 0.069 & 0.503 & 0.037 & 
                & 0.060 & 0.467 & 0.030 &
                & 0.154 & 0.458 & 0.075 \\
05:31$^2$       & 0.137 & 0.392 & 0.057 & 
                & 0.122 & 0.379 & 0.049 & 
                & 0.111 & 0.350 & 0.041 &
                & 0.275 & 0.401 & 0.118  \\
06:01           & 0.111 & 0.425 & 0.050 & 
                & 0.102 & 0.381 & 0.042 &  
                & 0.089 & 0.359 & 0.034 &
                & 0.222 & 0.402 & 0.095  \\
06:29           & 0.093 & 0.342 & 0.034 & 
                & 0.082 & 0.349 & 0.030 &  
                & 0.069 & 0.337 & 0.025 &
                & 0.177 & 0.416 & 0.078  \\
07:29           & 0.103 & 0.414 & 0.045 & 
                & 0.100 & 0.318 & 0.034 &  
                & 0.082 & 0.265 & 0.023 &
                & 0.204 & 0.387 & 0.084  \\
\noalign{\smallskip}
\hline
\noalign{\smallskip}
\multicolumn{8}{l}{$^1$Quiescent spectrum; $^2$Flare maximum }
\end{tabular}

\end{flushleft}
\end{table*}

\subsection{The flare in the UV} 

\begin{figure*}
{\psfig{figure=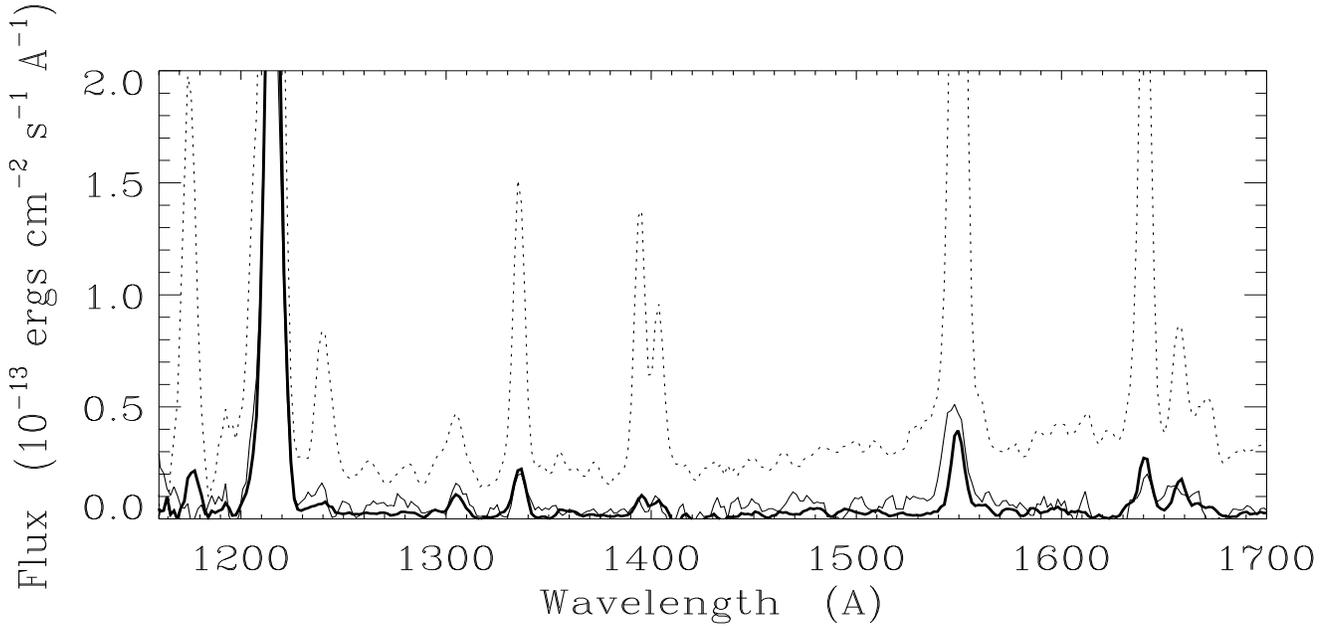,bbllx=43pt,bblly=400pt,bburx=554pt,bbury=680pt,height=8.4cm,width=17.5cm}}
\caption[ ]{
IUE SWP-lo spectra showing fluxes at earth (per \AA)
for the mean quiescent phase (average of 10 spectra; heavy solid), the
early impulsive phase (JD 2449343.568; thin solid) and
the near-maximum phase  (JD 2449343.783; dotted).  The spectra have been
smoothed by a 3 pixel running mean.
\label{fig:uv_spec_s3} }
\end{figure*}

\begin{figure}
{\psfig{figure=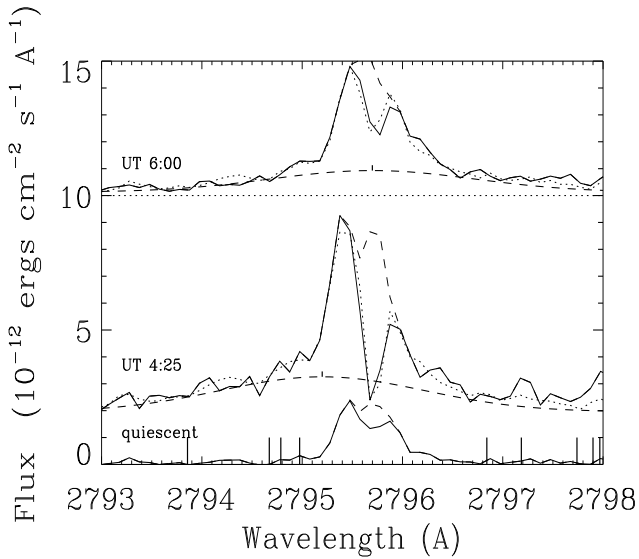,bbllx=102pt,bblly=374pt,bburx=548pt,bbury=719pt,height=7.4cm,width=8.4cm}}
\caption[ ]{IUE LWP-hi spectra of the Mg~{\sc ii} k line
showing fluxes at earth (per \AA)
for the mean quiescent phase (average of 8 spectra; solid), and
two spectra near flare peak (taken between the
two SWP spectra in Fig.~\ref{fig:uv_spec_s3}); the later spectrum
is offset by 10$^{-11}$ for clarity.
Spectra with the ISM absorption approximately removed (dashed)
and simple models combining ISM, enhanced quiescent, and broad
components (see text) are shown (dotted); blends are also
noted (thick vertical marks), as is the broad component peaks
(note the shift between the spectra).
\label{fig:lqmg2} }
\end{figure}

\begin{figure}
{\psfig{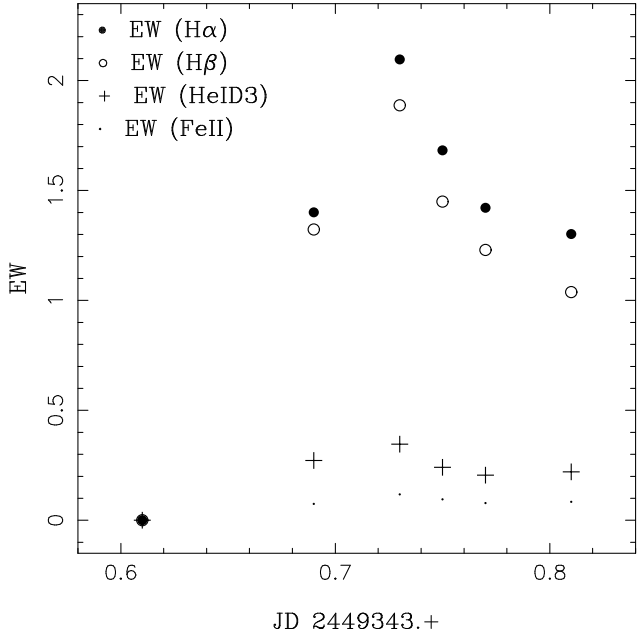}}
\caption[ ]{The change of EW of several optical chromospheric lines
 during the flare
\label{fig:lqhya_ews} }
\end{figure}

\begin{figure}
{\psfig{figure=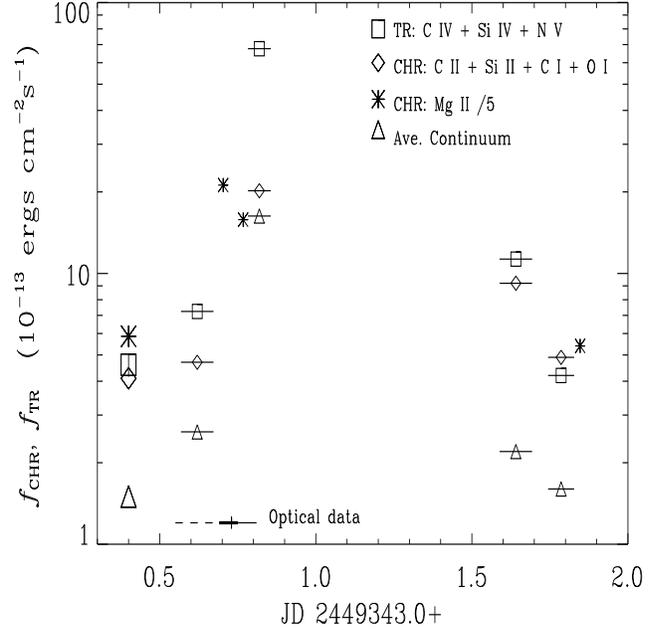,bbllx=64pt,bblly=369pt,bburx=546pt,bbury=706pt,height=8.4cm,width=8.4cm}}
\caption[ ]{Evolution of the combined IUE chromospheric
fluxes (at earth) of O~{\sc i} (1304~\AA), C~{\sc ii} (1335\AA), 
C~{\sc i} (1657~\AA), and Si~{\sc ii} (1810~\AA)  
(=$f_{\rm CHR}$, $\diamond$), 
chromospheric Mg~{\sc ii}$/5$ (2800~\AA; $\ast$), the 
combined transition region fluxes of N~{\sc v} (1240~\AA),
Si~{\sc iv} (1400\AA) and C~{\sc iv} (1550\AA) (=$f_{\rm TR}$, $\Box$),
and the UV continuum flux (average of the total $f_{\rm cont}$ in two 
50~\AA~bands centered at 1490~\AA~and 1595~\AA; $\triangle$). 
Horizontal solid lines through the data points indicate
the exposure durations.  The first time points (plotted arbitrarily at 0.4)
show the mean quiescent fluxes and their errors (vertical line).
The span of the optical observations is also indicated, with the line type
indicating the spectrum's appearance (dashed=quiescent,
heavy solid=impulsive/flare peak, thin solid = gradual); 
the optical maximum is marked with a tick. 
\label{fig:lqhya_uv} }
\end{figure}

Ultraviolet line and continuum fluxes are given in (Table \ref{tab:uv_fluxes}). 
We coadd selected strong chromospheric (CHR) and transition region (TR) 
line fluxes to improve statistics:  $f_{\rm TR}$ is the
sum of N~{\sc v} ($T_{\rm form} \sim 1.25\times10^5$ K), 
Si~{\sc iv} ($T_{\rm form} \sim 8\times10^4$ K) and C~{\sc iv} ($T_{\rm form}
\sim 10^{5}$ K), while $f_{\rm CHR}$ coadds O~{\sc i} and C~{\sc i} 
($T_{\rm form} \sim
7\times10^3$ K), C~{\sc ii} ($T_{\rm form} \sim 2\times10^4$ K), 
and Si~{\sc ii} ($T_{\rm form}\sim 1\times10^4$ K). 
Figure~\ref{fig:uv_spec_s3} shows the quiescent (an average of 10
exposures), impulsive (JD$_{\rm start}$ =  2449343.568; 1:37 UT), 
and just post-maximum (= ``UV peak"; 
JD$_{\rm start}$ =  2449343.783; 6:48 UT) UV spectra.
Figure~\ref{fig:lqmg2} depicts Mg~{\sc ii} spectra
between the SWP data in Figure~\ref{fig:uv_spec_s3}, plus an
average quiescent spectrum.
Figure~\ref{fig:lqhya_uv} shows the time variation of $f_{\rm TR}$
and $f_{\rm CHR}$ during the period of the optical flare.  We note that
while the UV ``impulsive" spectrum ends before the first flare-affected
optical spectrum (4:36 UT), 
it already shows noticeably enhanced (1.5$\times$quiescent) transition
region (TR) and continuum emission.  Thus the true ``impulsive" phase 
starts earlier than the optical data
indicates, with high $T_{\rm form}$ TR lines leading the evolution 
(as is typical).

Besides the lines measured in Table \ref{tab:uv_fluxes}, 
numerous weak lines are also present (typically in UV peak 
and/or quiescent spectra).  Most certain of these are
C~{\sc i} 1261, 1277, 1280, 1560~\AA, 
Si~{\sc iii} $\lambda$ 1295-98, 1892~\AA, 
S~{\sc i} 1820~\AA, 
Al~{\sc iii} 1855, 1863~\AA,
Fe~{\sc ii} 1611, 1633, 1671~\AA, and
O~{\sc iii} 1666~\AA.
The following are also detected with less certainty:
C~{\sc i} 1463, 1493~\AA, 
S~{\sc i} 1473, 1483, 1487, 1900~\AA, 
Si~{\sc i} 1258, 1267, 1565, 1574~\AA,
Si~{\sc ii} 1533\AA, Fe~{\sc ii} 1372, 1588, 1598~\AA,
S~{\sc iv} 1417~\AA,
O~{\sc iv} 1407~\AA, 
O~{\sc v} 1371~\AA, and
S~{\sc v} 1502~\AA.
Also of interest is the possible detection of the coronal 
1354.1~\AA\ Fe~{\sc xxi} line ($T_{\rm form} \approx 10^7$ K) in the
UV peak and (perhaps) in the quiescent spectrum.
Possible contributions from other nearby lines
(e.g., C~{\sc i} 1354.3~\AA) make 
this identification uncertain, though.


The strongest high $T_{\rm form}$ line, C~{\sc iv}, 
shows a strong blueward asymmetry in the impulsive spectrum 
(its centroid is shifted by $\approx -250$ km s$^{-1}$), 
perhaps foreshadowing the
blueshifts in the broad components of the optical Balmer and He~{\sc i} lines
seen later (see \S 3.5). Si~{\sc IV} and N~{\sc V}
(with similar $T_{\rm form}$) also appear to be blue-shifted,
though the shift is uncertain due 
to probable blends and the weakness these features.
Cooler strong lines (C~{\sc II}, He~{\sc II}) 
show no shift.  The lack of H$_2$ rules out the 1547~\AA~line as responsible,
but we note that numerous
features near C~{\sc IV} are coincident with (normally weak) Si~{\sc I} 
(1545, 1552, 1555-8, 1563~\AA), and C~{\sc I} (1542~\AA) lines.
Thus, the apparent C~{\sc IV} blueshift {\it may} be partly due to blends.

There are no SWP data during optical flare maximum, 
though two LWP spectra appear to show Mg~{\sc ii} peaking
at about the this time (Fig.~\ref{fig:lqmg2}).  The presence of ISM absorption, 
emission self-reversal,
and nearby blends complicates the interpretation of these features. 
We first constructed a ``quiescent" spectrum $F_{\rm Q}$ 
from the average of
eight LWP images which appeared uncontaminated by flares. 
We then removed the ISM absorption from $F_{\rm Q}$ by adding
a 2 pixel Gaussian at the center of the absorption feature.
We modeled the Mg~{\sc ii} lines in the flare very simply, with
$F_{\rm flare} = A F'_{\rm Q} + B G_{\rm B} - C G_{\rm ISM} +D$,
where $F'_{\rm Q}$ is the ISM-corrected quiescent spectrum,
$G_{\rm B}$ and $G_{\rm ISM}$ are broad and narrow (2 pixel) Gaussians,
and $A$, $B$, $C$, and $D$ are adjustable constants.
The width and central $\lambda$ of $F'_{\rm Q}$ and $G_{\rm B}$
were also allowed to vary; results are shown in Fig.~\ref{fig:lqmg2}.  
The advantage of using $F'_{\rm Q}$ as a template is
that we account for the intrinsic Mg~{\sc ii} shape and 
(partially) remove blends.

We find that the first Mg~{\sc ii} spectrum shows a 2.5$\times$
enhancement over quiescent in its narrow component, and a broad 
($\approx$250 km s$^{-1}$ FWHM) component with a flux
of $f_{\rm B} = 3.3\times10^{-12}$ ergs cm$^{-2}$ s$^{-1}$ 
at earth in Mg~{\sc ii} k.  The broad component in this spectrum
is Doppler shifted by $\approx -40$ km s$^{-1}$.
The continuum is also significantly enhanced (by at least 10$\times$), with 
$\langle f_{\rm con} \rangle \approx 2\times10^{-12}$
ergs cm$^{-2}$ s$^{-1}$ \AA$^{-1}$ at 2800~\AA.
The second spectrum, taken about 1.5 hours later,
showed both broad and narrow components reduced by a factor of
$\approx 1.5$, with the broad component now basically unshifted but
with a similar width (Fig.~\ref{fig:lqmg2}).
Thus, the Mg~{\sc ii} lines respond to the flare
much like the optical chromospheric features, initially exhibiting
enhanced emission with broad, blue-shifted   
components, which gradually drift to the red and weaken 
as the flare evolves.

The  next SWP spectrum (UV peak, 
coinciding the with early gradual phase in the optical
data) reveals $> 20 \times$ enhancements in the TR and $> 4 \times$
enhancements in the CHR.  Several high $T_{\rm form}$ lines
only weakly detected in the quiescent spectrum (C~{\sc iii} 1175~\AA,
N~{\sc v} 1240~\AA) are also greatly strengthened.  Several lines not listed 
in Table \ref{tab:uv_fluxes} appear in this spectrum; the more certain of these
include the C~{\sc i} complexes (1261~\AA, 1277~\AA, and 1280~\AA) and the
Si~{\sc iii} complex at 1295-1299~\AA, 
(combined fluxes $f \approx$ 1.3 and $0.9 \times 10^{-13}$ ergs 
cm$^{-2}$ s$^{-1}$),  O~{\sc iii} 1666~\AA, Fe~{\sc ii} 1671~\AA,
Al~{\sc iii} 1854+1862 \AA, and Si~{\sc iii} 1892~\AA\ (with $f \approx 
0.8, 1.3$, $2.4$, and $1.3\times 10^{-13}$ ergs cm$^{-2}$ s$^{-1}$, 
respectively).

The UV continuum is also noticably enhanced in this spectrum.
Machado \& H\'enoux (1982) suggested similar enhancements in solar flares
at $\lambda <$ 1524~\AA~and $\lambda <$ 1682~\AA  are due to are due bound-free
radiation from the ground and first excited levels of Si~{\sc i}
excited by UV line emission (primarily by C~{\sc ii} and
C~{\sc iv}, respectively).  Exploring this idea, 
Phillips, Bromage \& Doyle (1992) 
studied  the power per \AA~in 50 \AA~(relatively line free) bands centered at
1490~\AA~and 1595~\AA~in stellar flares, and found that they correlated
well with line power in C~{\sc ii} and C~{\sc iv}, respectively:
specifically, $\log P_{\rm cont}$(1490\AA) = 0.94 $\log P_{\rm C~{\sc ii}} + 
0.3$ and $\log P_{\rm cont}$(1595\AA) = 1.04 $\log P_{\rm C~{\sc iv}} - 2.9$.  
We have integrated the flux in similar bands ($f_{\rm cont}$ in 
Table \ref{tab:uv_fluxes}); note that the small $f_{\rm cont}$ in the
non-flare spectra may be due to the sum of many weak lines.
We find good agreement for the 1490~\AA~continuum, 
with $\log P_{\rm cont}$(1490~\AA) (= $f_{\rm cont}/50$\AA\ $\times 4 \pi d^2$, 
where $d$ is the stellar distance) within an average 0.10 dex of the prediction
for the impulsive and UV max spectra. 
Agreement is less good for $\log P_{\rm cont}$(1595~\AA), which is 
overestimated by an average of 0.31 dex. 
We see no Si~{\sc i} edges at 1524~\AA~or 1682~\AA, 
but this is perhaps not surprising
given the low resolution of IUE (Phillips et al.~1992).
In general, though, the far UV flare
continuum of LQ Hya seems consistent with the Si~{\sc i} recombination model.

If LQ Hya were very young 
and still had significant circumstellar material, one might expect X-ray 
excitation of the gas in such a strong flare.
On the suggestion of the referee, we explored whether any of the weak features
in the impulsive or UV peak spectrum might be due to excitation of
circumstellar H$_2$.  As there is no evidence in these spectra 
for the four strongest H$_2$
features seen in T Tau (1446, 1490, 1505, 1562~\AA; Brown, de M. Ferraz \& Jordan~1984), 
it seems unlikely
there is much nearby gas.  This makes it less likely that LQ Hya is
pre-main sequence, as suggested by Vilhu et al. (1991).

The UV fluxes do not return to their 
quiescent level until almost a day later (JD 2449344.744), though lack of
data for $\sim$0.8 day after the UV maximum makes the
interpretation ambiguous. The time 
between UV maximum and quiescence covers over half a 
rotation, making it difficult for the flare (if localised) to have
remained visible unless it was very near the pole, or in a very
extended ($>R_*$) loop.  The sharp drop in flux  from
344.589 to 344.744 could then be due in part to the flare's finally
disappearing over the limb.  Perhaps more likely, the enhancement seen on JD
2449344 might be due to a second flare and the fast decay a 
consequence of its lower energy ($E_{\rm flare} \propto t_{\rm decay}^2$;
Lee, Petrosian \& McTiernan 1993; Saar \& Bookbinder 1998).

\begin{table*}
\caption[]{UV continuum fluxes, chromospheric (CHR) and transition region 
(TR) lines fluxes
\label{tab:uv_fluxes}}
\begin{flushleft}
\begin{tabular}{lccccccccccccccc}
\hline
\noalign{\smallskip}
 JD$_{\rm start}$ & t$_{\rm exp}$ & 
\multicolumn{10}{c}{$f$~~~(10$^{-13}$ ergs cm$^{2}$ s$^{-1}$ at earth)$^1$} &
 $f_{\rm CHR}$ & $f_{\rm TR}$ & $f_{\rm cont}$ & $f_{\rm cont}$ \\  
 -2449000 & (s) & C {\sc iii} & N {\sc v} & O {\sc i} & C {\sc ii} & 
 Si {\sc iv} & C {\sc iv} & He {\sc ii} & C {\sc i} & Si {\sc ii} &
Mg {\sc ii}$^2$ & sum & sum & {\scriptsize (1490\AA)} & 
{\scriptsize (1595\AA)}  \\
\hline
\noalign{\smallskip}
%
342.742 & 2400 & ...  & ...  & ...  &  ... &  ...    
 & ... &  ... & ... & ... & 30/32 &  ... &  ... & ... & ...   \\
342.781$^{3}$ & 7200 & 0.8:$^{4}$  & 0.2:  & 0.8  &  1.6 &   1.4    
 & 3.3 &  2.0 & 1.1 & 1.5 & ... & 5.0 &  4.8 & 0.9: & 1.6:  \\
343.568$^{5}$ & 9000 & $<$0.4: & 0.6:  & 0.9  &  1.3 &  0.8:    
 & 5.8: &  1.2 & 1.3 & 1.2 & ... &  4.7 &  7.2: & 2.4: & 2.9:    \\
343.684 & 2400 & ...  & ...  & ...  &  ... &  ...    
 & ... &  ... & ... & ... & 106:/120: &  ... &  ... & ... & ...   \\
343.750 & 2400 & ...  & ...  & ...  &  ... &  ...    
 & ... &  ... & ... & ... & 78/83 &  ... &  ... & ... & ...   \\
343.783       & 6300 & 13.7   & 5.4  & 2.8  &  8.8 &  12.2 
 & 50:$^6$ & 18.3 & 4.5 & 4.1 & ... &20.2 & 68: &  14.7 & 18.0   \\
344.589       & 9000 & 3.8:   & 1.5:  & 1.3:  &  3.1 &   2.9 
 &   6.9 &  3.5 & 2.0 & 2.8 &  ... &9.2 &  11.3 & 1.6: & 2.9:  \\
344.744       & 7200 & 1.0: & 0.4:  & 0.4:  &  1.0 &  0.5 
 &   2.4 &  1.7 & 1.2 & 1.1 &  ... &4.9 &  4.2 & 1.6: & 1.6:  \\
344.833 & 2400 & ...  & ...  & ...  &  ... &  ...    
 & ... &  ... & ... & ... & 27/29 &  ... &  ... & ... & ...   \\
\noalign{\smallskip} 
\multicolumn{2}{l}{$<$quiescent$>$} 
                      & 1.2: & 0.4 & 0.7 & 1.5 &  1.1 
 &   3.1 &  1.7 & 1.0 & 0.9 &  29/31 & 4.1 &  4.6 & 1.5 & 1.5  \\
\noalign{\smallskip}
\hline
\noalign{\smallskip}
\end{tabular}
$^1$ lines at 1175\AA, 1239+43\AA, 1302+5+6\AA, 1335\AA, 1394+1403\AA, 
1548+51\AA, 1641\AA, 1657\AA, 1808+17\AA, 2796+2803\AA, respectively.

$^{2}$ uncorrected/corrected for ISM absorption~~~~~$^{3}$ not shown in Fig. 
10.~~~~~$^{4}$ colon = uncertain measurement~~~~~$^{5}$ noisy spectrum

$^{6}$ line saturated, $f > 38.0 \times 10^{-13}$ ergs cm$^{-2}$ s$^{-1}$, 
flux estimated from Gaussian fit to line wings and shoulder. 

\end{flushleft}
\end{table*}

\subsection{Estimate of energy released}

To estimate the flare energy released in the observed optical 
chromospheric lines, we converted the EW into absolute
surface fluxes and luminosities.
Since we have not observed the
entire flare, and are missing important lines (e.g., the saturated
Ly $\alpha$; Mg~{\sc ii}; He~{\sc i} 10830~\AA) 
our estimates are only lower limits to the total flare energy 
in chromospheric lines.
We have used the calibration of Hall (1996) to obtain the 
stellar continuum flux in the H$\alpha$ region 
as a function of (B~-~V) and then convert
the EW into absolute surface flux.
For the other lines, we have used the continuum flux at
H$\alpha$ corrected for the difference in the continuum 
F$_{\lambda 6563}$/F$_{\lambda}$, assuming F$_{\lambda}$ is given by a 
blackbody at T$_{eff}$ = 4900 K. The corresponding 
absolute fluxes at flare maximum
(erg cm$^{-2}$ s$^{-1}$), and total flux (ergs cm$^{-2}$) integrated over the
the observation interval ($\sim$ 3~h) are given in Table~\ref{tab:energy}.
We converted these fluxes into luminosities using
the radius R = 0.76 R$_\odot$ (Strassmeier et al.~1993).
We have not estimated the energy released in the UV chromospheric lines,
since, without the (saturated) Ly$\alpha$ and Mg~{\sc ii}
lines, the total $f_{\rm CHR}$ will be greatly underestimated.
We find that the estimated total energy released in optical chromospheric 
lines is $E_{\rm CHR} \geq$~5.7~10$^{33}$ ergs, indicating 
that this LQ Hya flare is more energetic in line emission
than an average flare on a dMe star, where
typically 10$^{28}$ ergs $\leq E_{\rm CHR} \leq$ 10$^{34}$ ergs
(see Hawley \& Pettersen 1991).

\begin{table*}
\caption[]{The flare energy released the 
chromospheric lines
\label{tab:energy}}
\begin{flushleft}
\begin{tabular}{lccccc}
\hline
\noalign{\smallskip}
 {Line} 
& F$_{\lambda}${\scriptsize (max)} & $\int$F$_{\lambda} dt$
& L$_{\lambda}${\scriptsize (max)} & $\int$L$_{\lambda} dt$ \\
& (10$^{6}$)                             & (10$^{10}$)   
& (10$^{29}$)                            & (10$^{33}$) \\  
& {\scriptsize (ergs cm$^{-2}$ s$^{-1}$)} & {\scriptsize (ergs cm$^{-2}$)}
& {\scriptsize (erg s$^{-1}$)}           & {\scriptsize (erg)}           \\ 
\hline
\noalign{\smallskip}
%
H$\beta$                  & 6.269 & 5.876 & 2.204 & 2.066 \\
Mg~{\sc i} b$_{3}$        & 0.200 & 0.210 & 0.070 & 0.074 \\
Fe~{\sc ii} $\lambda$5169 & 0.414 & 0.406 & 0.145 & 0.143 \\
Mg~{\sc i} b$_{2}$        & 0.172 & 0.174 & 0.062 & 0.061 \\
Mg~{\sc i} b$_{1}$        & 0.144 & 0.137 & 0.051 & 0.048 \\
He~{\sc i} D$_{3}$        & 1.270 & 1.234 & 0.447 & 0.434 \\
Na~{\sc i} D$_{2}$        & 0.217 & 0.253 & 0.076 & 0.089 \\
Na~{\sc i} D$_{1}$        & 0.239 & 0.278 & 0.084 & 0.098 \\
H$\alpha$                 & 7.510 & 7.242 & 2.641 & 2.546 \\
He~{\sc i} $\lambda$6678  & 0.415 & 0.412 & 0.146 & 0.145 \\
\noalign{\smallskip}
                          &       &       &       &       \\
Total lines               & 16.85 & 16.23 & 5.926 & 5.704 \\
%
\noalign{\smallskip}
\hline
\noalign{\smallskip}
\end{tabular}

\end{flushleft}
\end{table*}

\subsection{Line asymmetry}

Very broad wings have been found in Mg~{\sc ii} and in 
the subtracted profiles
of H$\alpha$, H$\beta$, He~{\sc i} D$_{3}$ and He~{\sc i} $\lambda$6678 lines,
and double Gaussian (narrow=N and broad=B) fits were used to model them.
The contribution of the broad component
to the total EW and FWHM of these lines reaches a maximum in the
impulsive phase and then decreases.
The line profiles are also asymmetric,
with the B component often shifted relative to the N.
In the impulsive phase we found a blue asymmetry in the lines -- 
the B component appears blue-shifted with respect to the N.
At the maximum of the flare $\Delta \lambda \approx 0$,
and during the gradual phase a red asymmetry is present,
with $\Delta \lambda$ increasing with time (see Tables 3-6).

Similar broad Balmer emission wings have been seen in 
dMe stars and chromospherically active binaries connected 
with flare events, with similar blue or red asymmetries in some cases
(Eason et al.~1992; Phillips et al.~1988; Doyle et al.~1988; 
Gunn et al.~1994a; Abdul-Aziz et al. 1995; Montes et al. 1996b, 1997; 
Montes \& Ramsey (1998); Abranin et al. 1998; Berdyugina et al. 1998)
and without obvious line asymmetries in other cases 
(Hawley \& Pettersen 1991).
The authors ruled out the possibility of pressure (Stark) effects as the cause
and conclude that the profiles are best explained
if these broad components and asymmetries are attributed
to plasma turbulence or mass motions in the flare.
In solar flares, most frequently, a red asymmetry is observed 
in chromospheric lines and interpreted as the result of 
chromospheric downward condensations (CDC) 
(Canfield et al. 1990 and references therein).
However, evidence of a blue asymmetry has been also 
reported (Heinzel et al. 1994) and blue and red asymmetries are observed 
simultaneously at different positions in a flaring 
region (Canfield et al. 1990), 
or at the same position but different times (Ji et al. 1994).
Recent line profile calculations 
(Gan, Rieger \& Fang 1993; Heinzel et al. 1994; Ding \& Fang 1997) 
show that a CDC can explain both the blue and red asymmetries. 
On the other hand, in stellar flares evidence of mass motions 
have also been reported, 
in particular a large enhancement in the far blue wings
of Balmer lines during the impulsive phase of a stellar
flare was interpreted 
as a high velocity mass ejection (Houdebine, Foing \& Rodon\'o 1990),
or high velocity chromospheric evaporation (Gunn et al. 1994b), 
and red asymetries in the wings of Balmer lines are reported
by Houdebine et al. (1993) as evidence of CDC.
Broadening and shifts in UV lines during stellar flares,
also interpreted as mass motions, have been
reported by Simon, Linsky \& Schiffer (1980), and Linsky et al. (1989).
Thus another possible explanation of the broad components and asymmetries we 
observed is mass motions in the flare - perhaps an explosive ejection/eruption
 (blueshift) followed by flows down the flaring loops (redshift).
Still, since
a CDC can also explain both the blue and red asymmetries observed in 
this stellar flare, it remains a distinct possibility as well.
%

\section{Conclusions}

We have detected a strong flare on LQ Hya in high resolution
optical spectra (4800 to 7000~\AA) and IUE SWP observations.
Such a strong flare is unusual in an early K star like LQ Hya. 
The flare started on 1993 December 22 between 
2:42 UT (quiescent spectrum) and 4:07 UT (end of first enhanced UV spectrum).
UV data suggest the impulsive phase began well before the first
optical sign at 04:36 UT.  The optical chromospheric lines
reached their maximum intensity $\approx$55 min later, by which time
the continuum enhancement had sharply decreased.  Thereafter, the optical line
emission slowly decreased in a gradual phase that lasted at least until
the last observation (07:29 UT).
Quiescent C~{\sc iv} flux levels were not recovered after $\approx$4~h UT 
on the following day (though a second flare or rotation of the flaring
region beyond the limb may have affected the results).

We detected an optical  continuum enhancement that increased toward
the blue ($\propto \lambda^{-1.35}$) and reached a maximum (36\%)
during the impulsive phase.  The UV continuum was 
enhanced by at least $\approx 10\times$ at 1500~\AA~and 2800~\AA. 

We analyse the lines by subtracting a quiescent LQ Hya spectrum or that of 
a modified inactive reference star.
The excess H$\alpha$ and H$\beta$ emission equivalent widths, 
in the observed - reference spectra, increase by a factor of 2.7 and 5.8 
at maximum, respectively, over the quiescent level. 
The H$\alpha$/H$\beta$ Balmer decrement is shallower during the flare, 
changing from 3.15 at the quiescent state
to 1.46 at the impulsive and maximum  phases of the flare.
We observe the He~{\sc i} D$_{3}$ line, a well known
diagnostic of flares, going into emission during the flare, 
reaching an EW = 0.346~\AA\ at the maximum.  We also observe excess emission
in He~{\sc i} lines at 4921.9, 5015.7, and  6678.1~\AA~in the (O-Q) spectra, 
and in other metal lines such as the Na~{\sc i} D$_{1}$ and D$_{2}$, the
Mg~{\sc i} b triplet and several Fe~{\sc i} and Fe~{\sc ii} lines.

The more intense lines,
H$\alpha$, H$\beta$, He~{\sc i} D$_{3}$ and He~{\sc i} $\lambda$6678,
exhibit very broad wings in the subtracted profiles.
Their profile are hence not well matched
with a single component  model; 
we have used a two Gaussian fit (narrow and broad) to characterize them.
For all these lines the contribution of the broad component
to the total EW of the line and their FWHM reach a maximum at the
impulsive phase and then decrease.
Moreover, the line profiles are asymmetric, appearing  
the broad component blue-shifted in the impulsive phase
and red-shifted during the gradual phase, with 
$\Delta \lambda$ between the components increasing with time.
Mg~{\sc ii} profiles respond similarly.
These broad components and asymmetries can be attributed 
to plasma turbulence or to upward and downward mass motions in the flare.
Similar blueshifts may be seen in the C~{\sc iv} line
during the impulsive phase of the flare.

Ultraviolet TR lines are enhanced by a factor of $> 20$, chromospheric lines
by a factor of $>4$ and the far UV continuum by a factor of $>10 \times$ over 
background in our UV spectrum just after optical flare maximum.  The 
continuua in our flare-affected UV spectra generally agree with a 
Si~{\sc i} recombination model.

We estimate the energy released during the flare for all the 
optical chromospheric lines at 
$\sim$ 5.7 10$^{33}$ ergs, indicating 
that this flare in LQ Hya is more energetic in line emission
than an average flare on a dMe star.

\section*{Acknowledgments}

This work has been partially supported by the Universidad Complutense de Madrid
and the Spanish Direcci\'{o}n General de Ense\~{n}anza Superior e Investigaci\'{o}n
Cient\'{\i}fica (DGESIC) under grant PB97-0259, by
NASA grants NAG5-1975 and NAGW-112, HST grant GO-5871.01-94A, 
and NSF grant AST-9528563.
YCU is supported by the Austrian Science Foundation grant S7302.
We thank the referee J.G. Doyle for helpful comments, and C.M. Johns-Krull
for useful discussions.


\end{document}